\documentclass[
preprintnumbers, 
superscriptaddress,
bibnotes,
amsmath,amssymb,
aps,
floatfix
]{revtex4-2}

\usepackage{dcolumn}
\usepackage{bm}

\usepackage{color}
\usepackage{footnote}
\usepackage{lineno}
\usepackage{amsmath}
\usepackage{comment}
\usepackage{siunitx}
\usepackage{mathtools}

\newcommand{\pr}[1]{\left(#1\right)}

\newcommand{\pg}[1]{\left[#1\right]}

\newcommand{\p}{^\prime}
\newcommand{\pp}{^{\prime \prime}}
\newcommand{\degC}{\mathrm{{}^\circ C}}
\newcommand{\CO}{(\textit{color online}) }

\newcommand{\one}{$\mathrm{I}$}
\newcommand{\two}{$\mathrm{I}\hspace{-1.2pt}\mathrm{I}$}
\newcommand{\three}{$\mathrm{I}\hspace{-1.2pt}\mathrm{I}\hspace{-1.2pt}\mathrm{I}$}
\newcommand{\four}{$\mathrm{I}\hspace{-1.2pt}\mathrm{V}$}

\renewcommand{\one}{A}
\renewcommand{\two}{B}
\renewcommand{\three}{C}
\renewcommand{\four}{D}

\begin{document}
\date{\today}
\title{Development of Neutron Interferometer using Multilayer Mirrors and Measurements of Neutron-Nuclear Scattering Length with Pulsed Neutron Source}

\author{Takuhiro~Fujiie}
\email{fujiie@phi.phys.nagoya-u.ac.jp}
\affiliation{Dept.~of~Phys. Nagoya University, Furocho Chikusa, Nagoya, 464-8602, Aichi, Japan}
\affiliation{RIKEN Center for Advanced Photonics, Hirosawa 2-1, Wako, 351-0198, Saitama, Japan}
 
\author{Masahiro~Hino}
\affiliation{Institute~for~Integrated~Radiation~and~Nuclear~Science, Kyoto~University, 2,~Asashiro-Nishi,~Kumatori,~Sennan-gun,~590-0494,~Osaka,~Japan}

\author{Takuya~Hosobata}
\affiliation{RIKEN Center for Advanced Photonics, Hirosawa 2-1, Wako, 351-0198, Saitama, Japan}

\author{Go~Ichikawa}
\affiliation{High Energy Accelerator Research Organization, Tokai, Ibaraki, 319-1106, Japan}
\affiliation{J-PARC Center, 2-4 Tokai, Ibaraki, 319-1195, Japan}

\author{Masaaki~Kitaguchi}
\affiliation{Dept.~of~Phys. Nagoya University, Furocho Chikusa, Nagoya, 464-8602, Aichi, Japan}
\affiliation{High Energy Accelerator Research Organization, Tokai, Ibaraki, 319-1106, Japan}
\affiliation{Kobayashi-Maskawa Institute, Nagoya University, Furocho Chikusa, Nagoya, 464-8602, Aichi, Japan}

\author{Kenji~Mishima}
\affiliation{High Energy Accelerator Research Organization, Tokai, Ibaraki, 319-1106, Japan}
\affiliation{J-PARC Center, 2-4 Tokai, Ibaraki, 319-1195, Japan}

\author{Yoshichika~Seki}
\affiliation{Institute of Multidisciplinary Research for Advanced Materials, Tohoku University, Katahira 2-1-1, Aoba-ku, Sendai 980-8577, Japan}

\author{Hirohiko~M.~Shimizu}
\affiliation{Dept.~of~Phys. Nagoya University, Furocho Chikusa, Nagoya, 464-8602, Aichi, Japan}
\affiliation{High Energy Accelerator Research Organization, Tokai, Ibaraki, 319-1106, Japan}

\author{Yutaka~Yamagata}
\affiliation{RIKEN Center for Advanced Photonics, Hirosawa 2-1, Wako, 351-0198, Saitama, Japan}


\begin{abstract}
This study entailed the successful deployment of a novel neutron interferometer that utilizes multilayer mirrors. 
The apparatus facilitates a precise evaluation of the wavelength dependence of interference fringes utilizing a pulsed neutron source. 
Our interferometer achieved an impressive precision of 0.02~rad within a 20~min of recording time.
Compared to systems using silicon crystals, the measurement sensitivity was maintained even when using a simplified disturbance suppressor.
By segregating beam paths entirely, we achieved successful measurements of neutron-nuclear scattering lengths across various samples. 
The values measured for Si, Al, and Ti were in agreement with those found in the literature, while V showed a disparity of 45\%. 
This discrepancy may be attributable to impurities encountered in previous investigations.
The accuracy of measurements can be enhanced further by mitigating systematic uncertainties that are associated with neutron wavelength, sample impurity, and thickness. 
This novel neutron interferometer enables us to measure fundamental parameters, such as the neutron-nuclear scattering length of materials, with a precision that surpasses that of conventional interferometers. 
\end{abstract}


\maketitle



Neutron interferometers, which are characterized by a neutron wave that is divided into two paths and then merged while sustaining its coherence\cite{Rauch2015-ai}, are used for the precise measurement of neutron-related interactions\cite{Sponar2021-cd}. 
Specifically, the Mach--Zehnder type neutron interferometer, which utilizes diffraction from a Si single crystal and was first demonstrated by Rauch in 1974\cite{Rauch1974-to}, has been broadly applied in a variety of physics experiments. 
These include measurements of neutron-nuclear scattering length\cite{Hammerschmied1981-xy, Rauch1985-ji, Ioffe1998-px}, demonstrations of classical physics with elementary particles\cite{Colella1975-tm, Werner1979-gm}, validations of quantum mechanics\cite{Rauch1978-vy, Cimmino1989-ix, Hasegawa2003-qz, Pushin2011-hw, Clark2015-md, Denkmayr2018-yo}, and investigations of exotic interactions as predicted by new physics\cite{Lemmel2015-yo, Li2016-qo}.
Recent proposals for experiments using neutron interferometers to probe new physics underscore their continuing significance\cite{Okawara2012-ik, Riedel2013-hd, Riedel2017-rd, Brun2019-ty, Rocha2021-kz, Iwaguchi2022-ha}. 


The sensitivity of the interactions measured is directly proportional to the neutron wavelength and the interaction length. 
However, neutron interferometers that use Si crystal are limited by the lattice constant in terms of the available neutron wavelengths and are also size-constrained by the dimensions of the ingot\cite{Rauch2015-ai}. 
Progress is being made in developing neutron interferometers capable of high-sensitivity measurements to overcome these limitations\cite{Pushin2009-vm, Sarenac2018-ra, Lemmel2022-qs}. 
Among them, Jamin-type neutron interferometers, which use neutron mirrors to reflect neutrons through an artificially fabricated multilayer structure to separate and recombine neutron paths\cite{Funahashi1996-du, Kitaguchi2004-rj, Seki2010-xs}, offer a significant advantage. 
The multilayer structure allows for the selection of neutron wavelengths based on specific design parameters\cite{Hino2009-kp}.


We introduce a novel neutron interferometer, integrating unpolarized neutron mirrors with a pulsed neutron source capable of identifying the wavelength from the time of flight (TOF) at J-PARC. 
The interferometer concurrently measures phase shifts at diverse wavelengths, bestowing a significant statistical advantage. 
It also enables tracking and rectification of time-varying disturbances beyond the pulse interval\cite{Yamamoto2021-ef}, thus circumventing the issues encountered by conventional systems using multilayer mirrors\cite{Saggu2016-qg}.
This paper accentuates that the developed interferometer achieved phase shift measurements with sensitivity comparable to systems employing Si crystals, even with simple disturbance suppressors. 
We also successfully conducted measurements of the neutron-nuclear scattering lengths for several nuclei. 
This breakthrough underscores a crucial advancement in neutron interferometry, enhancing sensitivity for fundamental physics involving neutrons.


The experiment was conducted at the low-divergence branch of BL05 (NOP) within the Materials and Life Science Experimental Facility (MLF) at J-PARC\cite{Nagamiya2012-xa, Nakajima2017-hb, Mishima2009-yx, Mishima2015-nm}. 
Pulsed neutrons, generated through the spallation reaction at a repetition frequency of $25~\unit{Hz}$, were moderated by a liquid hydrogen moderator, and directed to the experimental setup via a neutron mirror bender, a four-blade slit, and a vacuum guide tube. 
The beam power sustained during the experiments was 620~kW.

\begin{figure*}[ht]
\centering
\includegraphics[keepaspectratio,width=\linewidth]{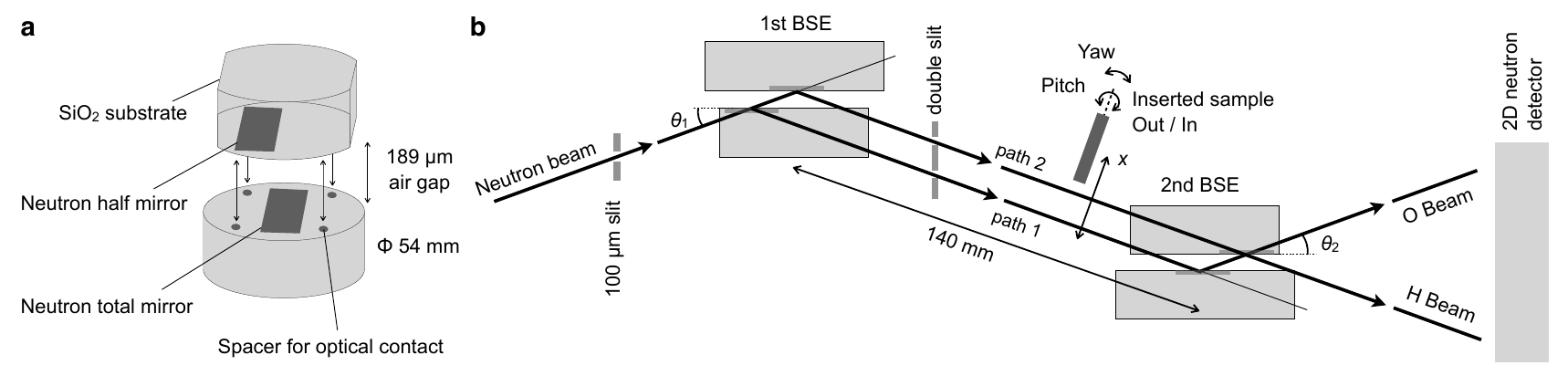}
\caption{
Illustration drawing of the design of our interferometer. 
(a) BSE with neutron mirror deposited. 
(b) Assembly of the interferometer, which is composed of two BSEs. 
Here, $\theta_1$ and $\theta_2$ signify the incident angles of the neutron wave onto the 1st and 2nd BSE, respectively. 
Beam size perpendicular to $x$ was collimated to 12 mm before 1st BSE.
}
\label{fig:setup}
\end{figure*}

Our interferometer comprised two Beam Splitting Etalons (BSEs)(Fig.~\ref{fig:setup}a) that split neutron waves into two parallel paths and later recombined them (Fig.~\ref{fig:setup}b).
The BSEs were designed with half and total reflection mirrors separated by an air gap of $D = 189~\unit{\micro m}$\cite{Seki2010-xs, Kitaguchi2003-zj}. 
The mirrors, deposited on a flat $\rm SiO_2$ substrate, were bonded by optical contact to maintain parallelism. 
The parallelism was preserved at approximately 30~nm, roughly in the range of typical transverse coherence length of the neutrons in this setup.
The neutron mirror, made from a Ni and Ti multilayer structure can reflect momentum transfers in the range of $0.232<Q<0.292~\unit{nm^{-1}}$\cite{Seki2011-bg}.
The BSEs were positioned at an incident angle, $\theta_{1}$ of $1.05~\unit{deg}$ with respect to the neutrons, with the centers of the two BSEs distanced $140~\unit{mm}$ apart. 
The neutron wave, having been divided into two paths, passed through a double slit situated 60~mm away from the center of the 1st BSE before being redirected towards the 2nd BSE.
This double-slit configuration consisted of two grooves, each $130$~\unit{\micro m} wide, separated by a center-to-center distance of $326 $~\unit{\micro m}.
This design was optimized based on the incident angle and refraction characteristics of the BSEs.
The interference pattern of the neutron waves was detected by a neutron detector equipped with time and two-dimensional position detection capabilities\cite{Hirota2005-cz}. 
The detector was placed 400~mm away from the center of the 2nd BSE.
To minimize external disturbances, the interferometer and sample insertion assembly were placed on an active vibration isolator and maintained at a stable temperature of $23 \degC$ inside a thermostatic chamber. 
Over 9.2~hours, the temperature's standard deviation remained at $0.035~\unit{K}$.

\begin{figure}[tbh]
    \centering
    \includegraphics[keepaspectratio,width=\linewidth]{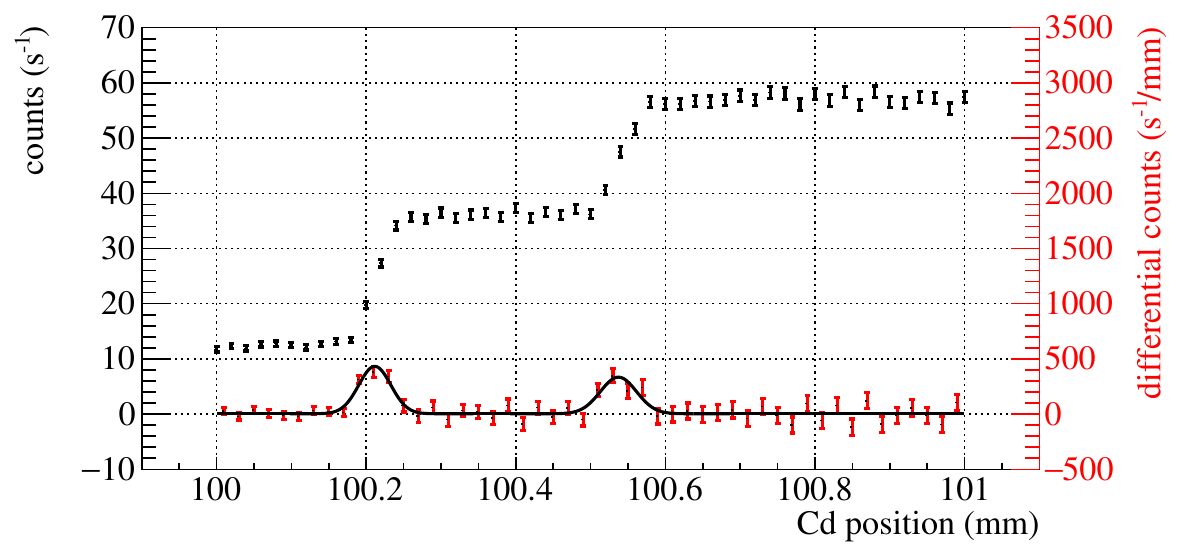} 
    \caption{\CO
    Overall neutron intensity of O and H beams in relation to the position of the Cd mask (black) and the corresponding differential intensity (red) that has been fit using a double Gaussian model.}
    \label{fig:step}
 \end{figure}

 
A Cd mask, 1~mm in thickness, was systematically traversed across the two beam paths located between the BSEs to quantify the beam profile. 
The mask sequentially shielded neutron waves by scanning in the $x$-direction using a sample insertion assembly. 
The neutron intensities pertaining to the O and H beams, along with their differentials, are shown in Fig.~\ref{fig:step}. 
The distinct staircase pattern in the neutron intensities indicates the complete bifurcation of neutrons into two separate paths by the 1st BSE and the double slit. 
A double Gaussian fitting applied to the derivative of the neutron intensity confirmed the path separation to be $326~\unit{\micro m}$, consistent with design specifications.

\begin{figure}[tbh]
    \centering   
    \includegraphics[keepaspectratio,width=\linewidth]{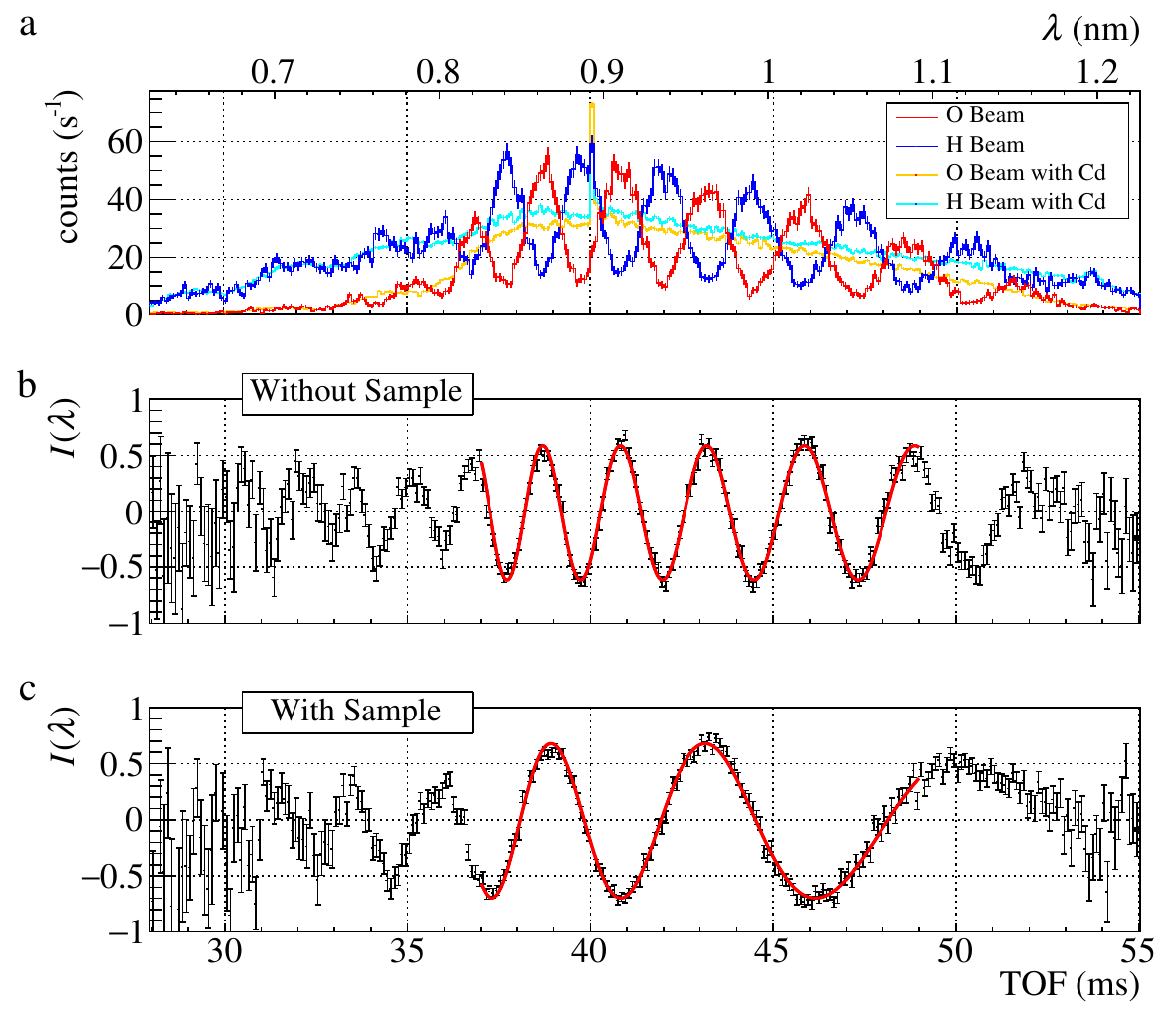} 
    \caption{\CO
    Time of flight (TOF) dependency of the measured interference fringes. 
    (a) TOF spectra of neutron intensity in the O and H beams. The spectra of pale color (orange and cyan) indicate the incoherent sum of both interferometer paths, obtained by blocking one or the other path with Cd and summing up both contributions.
    (b) Interference fringes obtained from the above spectrum.
    (c) Variation in the interference fringes prompted by the insertion of a Si sample. 
    The measurement time attributed to these fringes was 10~min. 
    For each spectrum, background readings were subtracted.
    }
    \label{fig:fringe}
 \end{figure}

The relative angle, $\delta\theta = \theta_2-\theta_1$, could be adjusted with a resolution of $6~\unit{\micro rad}$ using a micrometer affixed to the 2nd BSE. 
The parallel alignment of the 2nd BSE was achieved using an autocollimator with an accuracy of $3~\unit{\micro rad}$. 
After alignment, periodic intensity distributions were observed in the TOF region from $35~\unit{ms}$ to $50~\unit{ms}$ (refer to Fig.~\ref{fig:fringe}a). 
The absence of oscillation patterns when a single path was obstructed by a Cd mask indicates these oscillations were due to the interference of neutron waves from the two paths.


Interference fringes were calculated from the following equation:
\begin{equation}
    I(\lambda) = \frac{I_{\rm O}/I_{\rm O_{Cd}}-I_{\rm H}/I_{\rm H_{Cd}}}{I_{\rm O}/I_{\rm O_{Cd}}+I_{\rm H}/I_{\rm H_{Cd}}},
    \label{eq:exp_phase_shift}
\end{equation}
where $I_{\rm O}$ and $I_{\rm H}$ represent the measured TOF spectra of O and H beams, respectively, while $I_{\rm O_{Cd}}$ and $I_{\rm H_{Cd}}$ symbolize the spectra with a single path blocked by the Cd mask. 
Any prompt event stemming from proton collision with the mercury target within the $40<{\rm TOF}<40.1~\unit{ms}$ region was deliberately excluded from the analysis. 
The interference fringe, normalized by equation~\eqref{eq:exp_phase_shift}, is shown in Fig.~\ref{fig:fringe}b.
This interference fringe can be fitted by the following equation:
\begin{equation}
    I(\lambda) = A\cos\pr{\frac{P_{\rm L}}{\lambda} + P_{\rm R}\lambda - P_{\rm S}\lambda} + B,
    \label{eq:phase_shift}
\end{equation}
where $\lambda$ denotes the neutron wavelength (the derivation is provided in Appendix \one).
The initial term in the cosine function signifies the geometric optical length difference, $P_{\rm L} = 4\pi D \delta\theta$, between the two paths, where $D$ is the BSE's air gap. 
The second term, $P_{\rm R} = -4\pi D \delta\theta m U/(h^2\theta_1^2)$, denotes the path difference multiplied by the refractive index in $\rm SiO_2$ of the BSE. Here, $U$ represents the Fermi pseudopotential of $\rm SiO_2$, $m$ is the neutron mass, and $h$ is Planck's constant. 
The third term, $P_{\rm S} = Nb_{\rm c} t$, corresponds to the interaction with the inserted sample, where $N$ is the atomic number density, $b_{\rm c}$ is the neutron-nuclear scattering lengths, and $t$ is the thickness. 
The phase shifts, depicted as the third term, are directly linked to the sensitivity of the measured interactions. 
The visibility, represented by $A$, typically approximated 60\%. 
The effects of discrepancies between paths, such as differences in mirror reflectivity and neutron absorption by the sample, appear in the $B$ term. 
Typically it was about 3\%, but this does not affect the determination of the scattering length.
The fitting region was identified as the TOF range from $37~\unit{ms}$ to $49~\unit{ms}$, where the interference fringes were clearly discernible.


Fig.~\ref{fig:fringe}c shows the interference fringe produced upon inserting a 0.3~mm thick Si sample into one of the paths. 
A significant phase change occurred from the configuration without the sample. 
The uncertainty in the measured phase shift caused by the sample was obtained to be 0.02~rad per 20~min.
This sensitivity exceeded that measured by NIST (0.31~rad per min)\cite{Saggu2016-qg} by a factor of three and was comparable to that measured by ILL (0.08~rad per min)\cite{Hasegawa2003-qz}.
Contrary to the Si interferometer's requirement for a 5~mK temperature accuracy\cite{Pushin2015-nx}, which is an order of magnitude more stringent, the precision in phase determination is improved.
The time fluctuation of $P_{\rm S}$ was $3\times10^{-3}$~rad over a 20-min period, a value four orders of magnitude smaller than the phase shift caused by the Si sample, approximately $60$~rad. 
These results validate the interferometer's effectiveness within a simple thermostatic chamber, indicating its robust stability during extended measurement periods.

\begin{figure}[htbp]
\centering
\includegraphics[keepaspectratio,width=\linewidth]{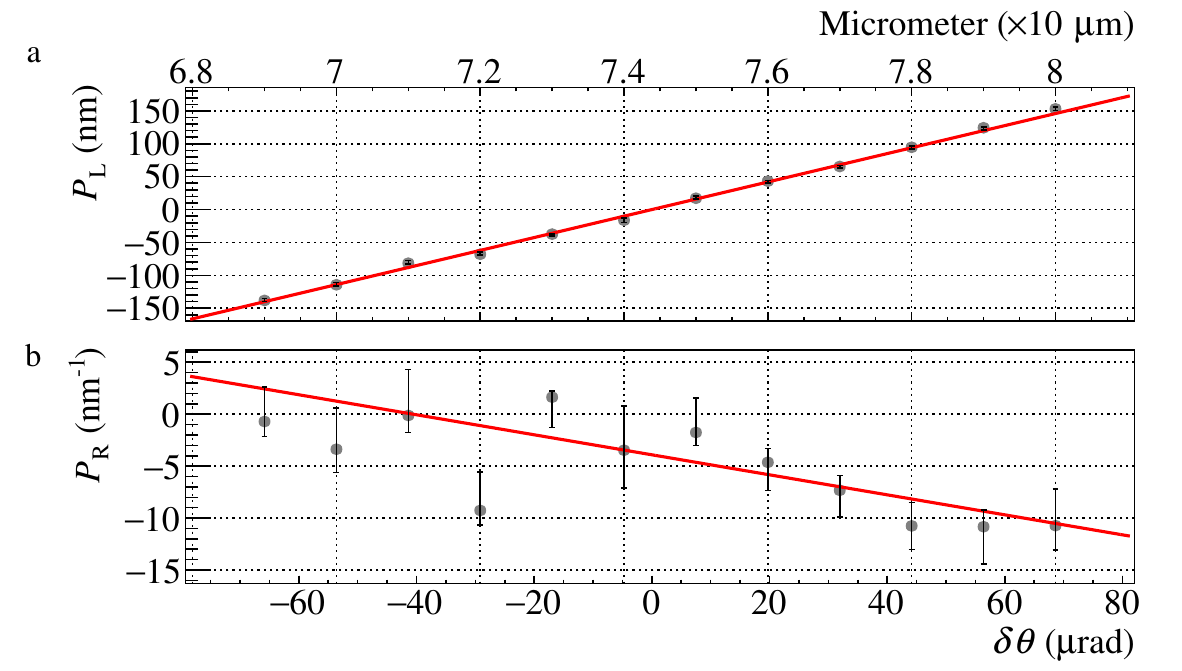}
\caption{
Relationship of $\delta\theta$ with (a) $P_{\rm L}$ and (b) $P_{\rm R}$, characterized by a linear function. 
The conversion factor between the micrometer value and $\delta \theta$ is established as $12.22~\unit{\micro rad/\micro m}$. 
The baseline value for $\delta \theta$ was determined through the fitting process applied to $P_{\rm L}$.
}
\label{fig:Si_p0}
\end{figure}

The phase of the fringes can be modified by adjusting $\delta\theta$ using the micrometer affixed to the 2nd BSE. 
The alteration in $P_{\rm L}$ relative to $\delta\theta$ was recorded as $P_{\rm L}/\delta\theta = 2.13 \pm 0.01~\unit{nm/\micro rad}$ (Fig.~\ref{fig:Si_p0}a), approximately 10\% less than the theoretical value of $2.38~\unit{nm/\micro rad}$. 
The change in $P_{\rm R}$ was estimated to be $P_{\rm R}/\delta\theta = -0.096 \pm 0.020~\unit{nm^{-1}/\micro rad}$ (Fig.~\ref{fig:Si_p0}b), nearly a quarter of the theoretical value of $-0.39~\unit{nm^{-1}/\micro rad}$. 
These deviations are speculated to arise from geometric inaccuracies in the BSEs or potential misalignment of the micrometer. 
Importantly, these effects can be compensated by calculating phase shifts in the presence and absence of the sample.


To demonstrate our interferometer's capabilities, we conducted experiments to determine neutron-nucleus scattering lengths. 
We selected Si and Al as representative nuclei, Ti and V as nuclei with negative $b_{\rm c}$ values, and V-Ni alloy as nuclei with near-zero $b_{\rm c}$. 
The V-Ni alloy is frequently employed in neutron scattering experiments as sample containers to prevent Bragg scatterings\cite{Sakurai2017-sq}. 
For accurate results, samples were precisely adjusted to minimize phase shifts in the interference fringes using three-axis stages (Fig.~\ref{fig:setup}b). 
Measurements of $P_{\rm S}$ were conducted with and without the sample for 5~min each to eliminate disturbances. 
This process was repeated over several hours for each sample (see Appendix~\two~for further details).


The obtained $b_{\rm c}$ values, along with their systematic and statistical uncertainties and the sample thicknesses, are listed in Table~\ref{table:result}.
The systematic uncertainties considered are detailed in Appendix~\three. 
The values for Si, Al, and Ti were consistent with the literature values, accurate to within 2.2\%. 
However, the results were generally smaller compared to the literature values, potentially due to systematic discrepancies between the TOF method-derived wavelengths and the true values\cite{noauthor_undated-kz}. 
Notably, the statistical uncertainty was two orders of magnitude less than the systematic uncertainty. 
Future work to enhance $b_{\rm c}$ determination accuracy should prioritize minimizing systematic uncertainties arising from sample conditions.


In contrast to the aforementioned samples, the results for the V and V-Ni alloy samples deviated significantly from their respective literature values. 
We conducted elemental analyses on the V and V-Ni alloy samples used in the measurements to investigate the underlying cause of this discrepancy. 
However, no significant impurities were identified. 
A comprehensive summary of all measured impurities can be found in Appendix~\four. 
To eliminate uncertainty arising from the neutron wavelength of the pulse source, we determined the $b_{\rm c}$ of V relative to Si using both V and V-Ni alloy samples.
The effects of measured impurities were also eliminated. 
The derived $b_{\rm c}$ values for V were $-0.555 \pm 0.003$ fm from the V sample and $-0.559 \pm 0.005$ fm from the V-Ni alloy sample. 
We used the $b_{\rm c}$ values of Si as $4.1491 \pm 0.0010$~fm and Ni as $10.3 \pm 0.1$~fm\cite{Varley_F1992-vf}. 
Both values of V were found to be in agreement, but deviated by 45\% from the NIST recommended value of $-0.3824\pm 0.0012~\unit{fm}$\cite{Bauspiess1978-ej}. 
Supporting this finding, a previous report suggested that the minimization of the Bragg peak required a higher amount of Ni in the V-Ni alloy compared to the NIST database\cite{Sakurai2017-sq}.


In conclusion, we have developed a neutron interferometer utilizing multilayer mirrors. 
The successful operation of our interferometer with pulsed neutrons at J-PARC yielded a visibility of 60\%.
Significantly, our interferometer demonstrated exceptional precision in phase determination, achieving an accuracy of 0.02~rad within a 20-min interval. 
Furthermore, our interferometer utilized a neutron wavelength of approximately  0.9~nm, providing enhanced sensitivity for specific applications compared to Si interferometers using 0.44~nm. 
The phase shift over time measured in our interferometer was four orders of magnitude smaller than the phase shift induced by a 0.3~mm thick Si sample.
Our interferometer demonstrates robustness against fluctuations, making it highly reliable. 
By introducing samples into one of the paths, we measured the $b_{\rm c}$ values for Si, Al, Ti, V, and V-Ni alloy. 
These measurements exhibited agreement with literature values within an accuracy of 2.2\%, except for vanadium, which requires further investigation to determine if impurities in the V sample contributed to the observed discrepancy.
These findings indicate that the phase shift introduced by the inserted samples was accurately captured. 


The statistical uncertainty associated with the measured phase shift was significantly smaller, up to two orders of magnitude, than the systematic uncertainty. 
This emphasizes the potential for highly sensitive measurements. 
To mitigate uncertainty in wavelength determination, two approaches can be employed: conducting relative measurements to Si or performing precise wavelength measurements. 
Furthermore, enhancing the sensitivity of $b_{\rm c}$ can be achieved by using thicker samples with reduced impurities. 
The systematic uncertainty arising from non-uniform sample thickness can be addressed through ultra-high precision machining technology\cite{Hosobata2017-bm, Hosobata2019-cz}.


Replacing the current multilayers with supermirrors and reflecting neutrons within the optimal wavelength range of 0.2--0.8~nm at J-PARC BL05 can increase statistics by a factor of 20.
This improvement would render the two-pathway separation interferometer nearly as sensitive as grating interferometers\cite{Sarenac2018-ra}.
Expanding the BSE's air gap by using a thicker spacer or by independently controlling each mirror could enable the measurement of various interactions, such as the scattering length of gas samples requiring containers of sufficient size\cite{Huber2014-lw, Lemmel2015-yo, Li2016-qo}.
The utilization of a high-sensitivity neutron interferometer opens up possibilities for a wide range of new physics search experiments\cite{Lemmel2015-yo, Li2016-qo, Rocha2021-kz, Okawara2012-ik, Riedel2013-hd, Riedel2017-rd, Brun2019-ty, Iwaguchi2022-ha}. 
The findings of this study represent the initial step towards realizing these experiments.


\begin{acknowledgments}
This research was supported by the JSPS KAKENHI Grant Number 21H01092 and the RIKEN Junior Research Associate Program. 
Financial support for this work was provided by JST SPRING, Grant Number JPMJSP2125. Takuhiro Fujiie would like to express gratitude to the ``Interdisciplinary Frontier Next-Generation Researcher Program of the Tokai Higher Education and Research System'' for their support. 
The neutron experiment conducted at the Materials and Life Science Experimental Facility of J-PARC was carried out under user programs (Proposal No.~2020A0226, 2020B0222, 2021B0109, and 2022A0116) and the S-type project of KEK (Proposal No.~2019S03). 
Finally, we extend our appreciation to the Advanced Manufacturing Support Team and Materials Characterization Support Team at RIKEN, as well as Prof.~Tatsushi~Shima, Dr.~Takuya~Okudaira, Shiori~Kawamura, and Rintaro~Nakabe for their invaluable support throughout the research process.
\end{acknowledgments}


\clearpage

\begin{table*}[ht]
\caption{
Each sample yielded measured $b_{\rm c}$. 
Reference values of $b_{\rm c}^{\rm ref}$ were taken from Sears (1992)\cite{Varley_F1992-vf}.}
\label{table:result}
\begin{ruledtabular}
\begin{tabular}{crrrrrr}
sample & $t$ (mm) & Purity (wt\%) & Stat. Unc. & Sys. Unc. & $b_{\rm c}$ (fm) & $b_{\rm c}$/$b_{\rm c}^{\rm ref}$ \\ \colrule

Si\footnote[1]{N-type wafer.} & $0.287 \pm 0.001$ & $99.999+$ & $9.2\times 10^{-5}$ & $6.7\times 10^{-3}$ & $4.060 \pm 0.027$ & $0.978 \pm 0.007$ \\
Ti & $0.100 \pm 0.002$ & $99.5+$ & $4.4\times 10^{-4}$ & $1.8\times 10^{-2}$ & $-3.477 \pm 0.062$ & $1.011 \pm 0.018$ \\
 & $0.204 \pm 0.004$ & $99.5+$ & $2.0\times 10^{-4}$ & $1.9\times 10^{-2}$ & $-3.386 \pm 0.064$ & $0.985 \pm 0.019$ \\
Al & $0.103 \pm 0.001$ & $99+$ & $6.0\times 10^{-4}$ & $1.5\times 10^{-2}$ & $3.408 \pm 0.050$ & $0.988 \pm 0.015$ \\
 & $0.287 \pm 0.002$ & $99+$ & $5.3\times 10^{-5}$ & $7.8\times 10^{-3}$ & $3.423 \pm 0.027$ & $0.992 \pm 0.008$ \\
 & $0.961 \pm 0.001$ & $99+$ & $4.1\times 10^{-5}$ & $5.7\times 10^{-3}$ & $3.466 \pm 0.020$ & $1.005 \pm 0.006$ \\
V & $0.316 \pm 0.001$ & $99.7+$\footnote[2]{The contaminations are summarized in Appendix \four.} & $3.9\times 10^{-4}$ & $6.8\times 10^{-3}$ & $-0.522 \pm 0.004$ & $1.364 \pm 0.010$ \\
 & $0.314 \pm 0.002$ & $99.7+$\footnotemark[2] & $6.1\times 10^{-4}$ & $7.2\times 10^{-3}$ & $-0.520 \pm 0.004$ & $1.361 \pm 0.011$ \\
V Averaged & --- & --- & --- & --- & $-0.521 \pm 0.003$ & $1.363 \pm 0.008$ \\
V-Ni alloy\footnote[3]{Fabricated by TAIYO KOKO Co., Ltd.}& $0.315 \pm 0.006$ & $\rm V:Ni=94.582:5.32$\footnotemark[2]  & $3.1\times 10^{-3}$ & $1.8\times 10^{-2}$ & $-0.062 \pm 0.001$ & $-0.528 \pm 0.010$ \\
\end{tabular}
\end{ruledtabular}
\end{table*}


\clearpage
\appendix

%
%
%
\section{Phase shifts by the geometric optical path}
\label{sec:phase_shift}
\begin{figure}[htbp]
\centering
\includegraphics[keepaspectratio,width=0.5\textwidth]{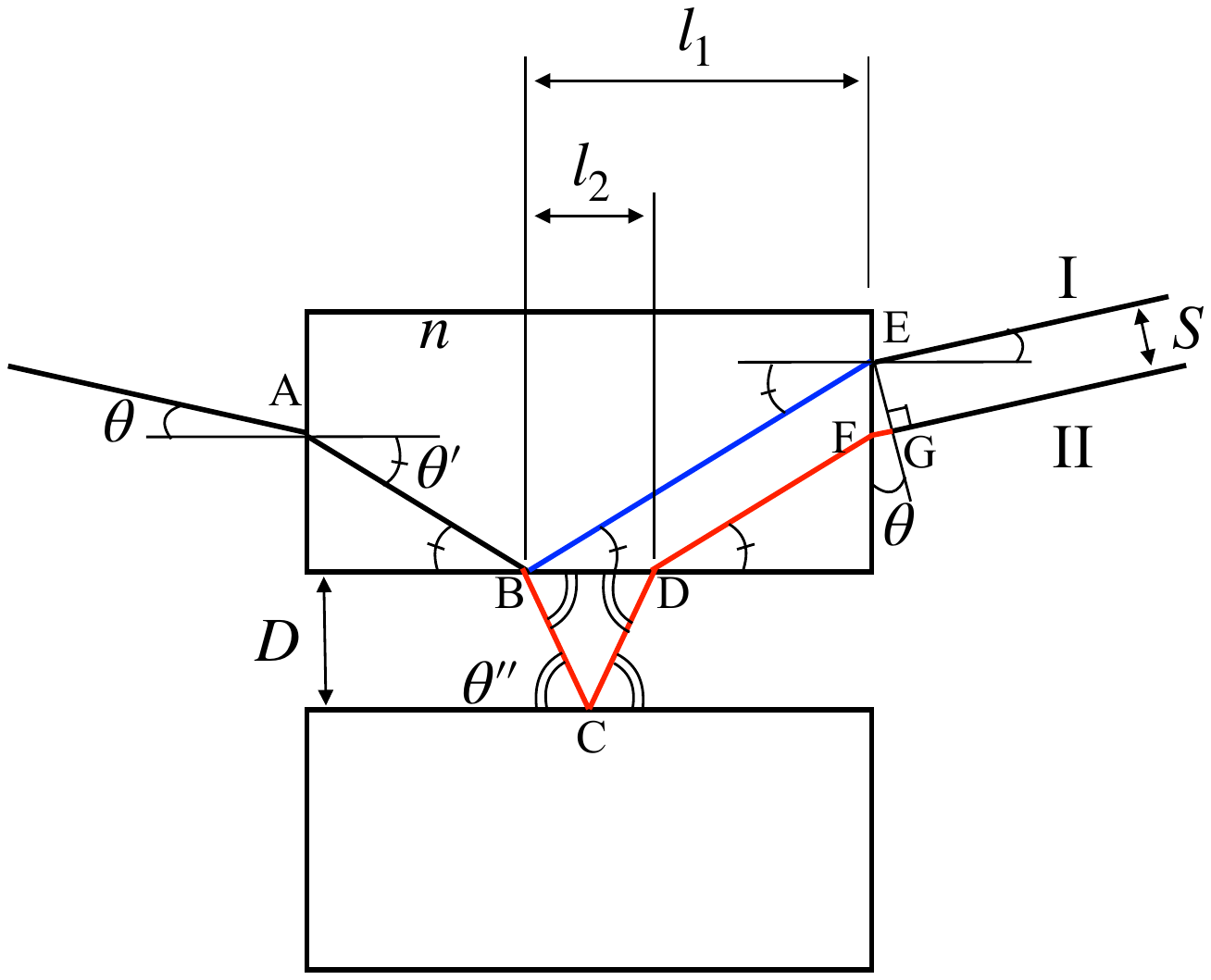}
\caption{Characteristics of the 1st BSE.
The neutron beam incidents from left to right.}
\label{fig:etalon_detail}
\end{figure}

Here, we provide a formal description of the variation in the optical path length of the multilayer-type interferometer. 
The diagram illustrating the paths of the neutrons in the first Beam Splitting Etalon (BSE) is presented in Figure~\ref{fig:etalon_detail}.
where $D=189~\unit{\micro m}$ represents the air gap of the BSE, $\theta=1.05~\unit{\deg}$ denotes the incident angle of neutrons, and $n$ corresponds to the refractive index of the glass substrate. The angles $\theta\p$ and $\theta\pp$ are considered as the angles accounting for refraction.


It is assumed that the two glass substrates of the BSE are parallel, and the sides and bottom of the glass substrate are perpendicular. 
The optical lengths $L$ of path $\mathrm{I}$ (B$\to$E) and path $\mathrm{I}\hspace{-1.2pt}\mathrm{I}$ (B$\to$C$\to$D$\to$F$\to$G) can be expressed as follows:
\begin{eqnarray}
  L_{\mathrm{I}}&=&\frac{n l_{1}}{\cos \theta\p}, \\
  L_{\mathrm{I}\hspace{-1.2pt}\mathrm{I}}&=&\frac{2 D}{\sin \theta\pp} +\frac{n\pr{l_{1}-l_{2}}}{\cos \theta\p} \notag \\ & & +\pr{l_{1} \tan \theta\p-\pr{l_{1}-l_{2}} \tan \theta\p} \sin \theta. \label{eq:path2}
\end{eqnarray}
The refraction index $n$ can be written as
\begin{eqnarray}
n &=& \sqrt{1-\frac{U}{E_n}} \notag \\ 
&\simeq& 1 - \frac{\lambda^2 N b_{\rm c}}{2\pi} \notag \\
&=& 1 - \frac{m U \lambda^2}{h^2},
\end{eqnarray}
where $U$ is the Fermi pesudopotential, $E_{\rm n}$ is the neutron energy, $N$ is the atomic density, $b_{\rm c}$ is the neutron-nuclear scattering length, $m$ is the neutron mass, and $h$ is the Plank constant.
Note that the absorption was neglected.
The relationship between the refraction angle and $n$ can be expressed using Snell's law as follows:
\begin{eqnarray}
\sin \theta &=n \sin \theta\p, \label{eq:snell01}\\
n \cos \theta\p &=\cos \theta\pp.\label{eq:snell02}
\end{eqnarray}
From the geometric conditions, the following equations hold:
\begin{eqnarray}
  1-\sin ^{2} \theta\pp &=&n^{2}\pr{1-\sin ^{2} \theta\p} \notag \\
  &=&n^{2}-\sin ^{2} \theta, \label{eq:geo03}\\
  \sin ^{2} \theta\pp &=&\pr{1-n^{2}}+\sin ^{2} \theta, \label{eq:geo04}\\
  \frac{\tan \theta\p}{\tan \theta\pp} &=&\frac{\sin \theta}{\sin \theta\pp}.\label{eq:geo05}
\end{eqnarray}
Using Eqs.~\eqref{eq:snell01}--\eqref{eq:geo05}, the Eq.~\eqref{eq:path2} is written as
\begin{eqnarray}
L_{\mathrm{I}\hspace{-1.2pt}\mathrm{I}}=\frac{2 D}{\sin \theta\pp}+\frac{n l_{1}}{\cos \theta\p}-\frac{2 n^{2} D}{\sin \theta\pp}+2 D \frac{\tan \theta\p}{\tan \theta\pp} \sin \theta.
\end{eqnarray}
Consequently, the optical path difference at the 1st BSE can be described as
\begin{eqnarray}
  \Delta L_{\text{1st BSE}} &=& L_{\mathrm{I}\hspace{-1.2pt}\mathrm{I}} - L_{\mathrm{I}} \notag \\
  &=& 2D\sqrt{(1-n^2) + \sin^2\theta} \notag \\
  &=& 2D\sqrt{1-n^2+\sin^2\theta}.
  \label{eq:deltaL}
\end{eqnarray}
By defining the normalized refraction index $\delta n$ as
\begin{eqnarray}
\delta n =1-n \simeq \frac{m U \lambda^2}{h^2},
\label{eq:refraction_index}
\end{eqnarray}
the Eq.~\eqref{eq:deltaL} can be written using $n^2 \simeq 1 - 2\delta n$, and applying an approximation of $\delta n/\sin^2\theta \ll 1$; then
\begin{eqnarray}
    \Delta L_\text{1st BSE} &\simeq& 2D\sqrt{\sin^2\theta + 2\delta n} \notag \\
    &\simeq& 2D\sin\theta\pr{1+\frac{\delta n}{\sin^2\theta}}.
\end{eqnarray}

Since the path difference for the 2nd BSE is the same as for the 1st BSE, the overall path difference in the interferometer is written as
\begin{eqnarray}
\Delta L &=& \Delta L_\text{2nd BSE} - \Delta L_\text{1st BSE} \notag \\
&\simeq& 2D\pg{\sin\theta_2\pr{1+\frac{\delta n}{\sin^2\theta_2}} -\sin\theta_1\pr{1+\frac{\delta n}{\sin^2\theta_1}}}, \notag\\
    \label{eq:deltaL_approx}
\end{eqnarray}
where the $\theta_1$ is the incident angle for the 1st BSE, $\theta_2$ is the incident angle for the 2nd BSE.
The air gaps of the two BSEs were assumed to be the same.
By denoting $\theta_2 = \theta_1 + \delta\theta$, $\theta_1 = \theta$, and assuming $\sin\theta \simeq \theta$, Eq.~\eqref{eq:deltaL_approx} can be written as
\begin{eqnarray}
    \Delta L \simeq 2D\delta\theta\pr{1 - \frac{\delta n}{\theta^2}}.
    \label{eq:deltaL_approx01}
\end{eqnarray}
The phase shifts due to the path difference are 
\begin{eqnarray}
\Delta \phi = \frac{2\pi}{\lambda}\Delta L.
\label{eq:phase_shift_origin}
\end{eqnarray}
From these, the phase shifts obtained in this experiment can be written as
\begin{eqnarray}
\Delta \phi \simeq 2\pi \frac{2D\delta\theta}{\lambda}\pr{1 - \frac{\delta n}{\theta^2}},
\label{eq:app_phase_shift}
\end{eqnarray}
which can be written using Eq.~\eqref{eq:refraction_index} as: 
\begin{eqnarray}
\Delta \phi &\simeq& 
\frac{4\pi D\delta\theta}{\lambda} - \frac{4\pi D \delta \theta mU}{h^2\theta{ }^{2}} \lambda.
\end{eqnarray}
Finally, the phase shift resulting from the difference in the optical path can be expressed using two terms.
One term is inversely proportional to the wavelength, while the other term is directly proportional to the wavelength.
These terms correspond to $P_{\rm L}$ and $P_{\rm R}$ in Eq.~(2).


\section{Data Analysis Process}
\label{sec:fitting}

\begin{figure}[htbp]
  \centering
  \includegraphics[keepaspectratio,width=\linewidth]
    {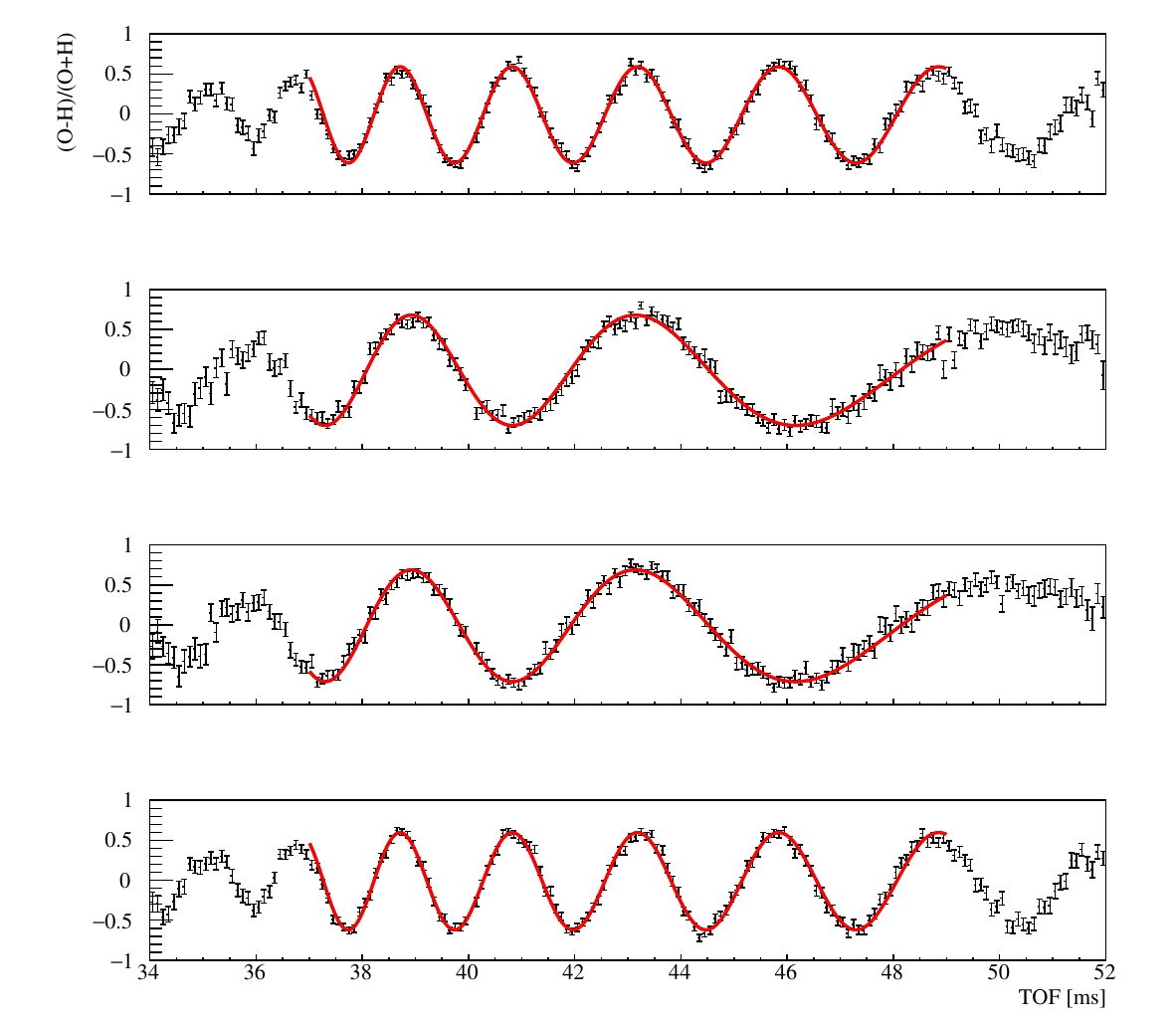}
    \caption{A typical interference fringe of one dataset. 
    The inserted sample was 0.3 mm thick Si.
    These fringes are the ``sample out-in-in-out'' from the top to the bottom.}
\label{fig:fringe_dataset}
\end{figure}

Interference fringe measurements were conducted in a sequence of a total of 20 minutes. 
In each set, the measurement sequence followed an ``out-in-in-out'' pattern, with a 5-minute interval between consecutive measurements. 
A global fitting analysis was then performed on the combined dataset consisting of four interference fringes.
This procedure effectively prevents substantial shifts in the fitting results caused by even the most minor disturbances.
The equation used for global fitting is as follows:
\begin{align}
f_0(\lambda) &= A_0\cos\pr{\frac{P_{\rm L}}{\lambda} + P_{\rm R}\lambda} + B_0 \\
f_1(\lambda) &= A_1\cos\pr{\frac{P_{\rm L}}{\lambda} + P_{\rm R}\lambda + P_\mathrm{S}\lambda} + B_1\\
f_2(\lambda) &= A_2\cos\pr{\frac{P_{\rm L}}{\lambda} + P_{\rm R}\lambda + P_\mathrm{S}\lambda} + B_2\\
f_3(\lambda) &= A_3\cos\pr{\frac{P_{\rm L}}{\lambda} + P_{\rm R}\lambda} + B_3,
\end{align}
where the $P_{\rm L}$ represents the optical path difference between two paths, $P_{\rm R}$ is the optical path times refraction index of $\rm SiO_2$, and $P_{\rm S}$ donated by the inserted sample (see main text).
In the four fittings, $P_{\rm L}$ and $P_{\rm R}$ were common parameters because the phase shifts due to geometrical optics do not change.
The $A$ and $B$ were independent parameters.
The fitting algorithm used was Migrad provided by CERN ROOT6.
The parameter with the smallest sum of $\chi^2/\mathrm{NDF}$ for the four fittings was considered the best fitting, where the NDF represents the number of degrees of freedom.
A typical dataset measured is shown in Fig.~\ref{fig:fringe_dataset}, along with the fitting functions.

The effect of the disturbance is observed as a slight change in $P_{\rm L}$ and $P_{\rm R}$, which appears in the $\chi^2/\mathrm{NDF}$ in the fit result. 
To account for this effect, The errors obtained by the fit were multiplied by $S = \sqrt{\chi^2/\mathrm{NDF}}$ when $S$ is greater than 1.


\section{Systematic Uncertainties}
\label{sec:systematics}
In the consideration of systematic uncertainties associated with the neutron-nuclear scattering length, the following factors were evaluated. 
The thickness of each sample was determined via the use of a digimatic micrometer of Mitutoyo Corporation (MDC-25SB).
A series of measurements were taken, and the mean and standard deviation of the measurements were used as the thickness of the sample.

\begin{figure}[htbp]
  \centering
  \includegraphics[keepaspectratio,width=\linewidth]{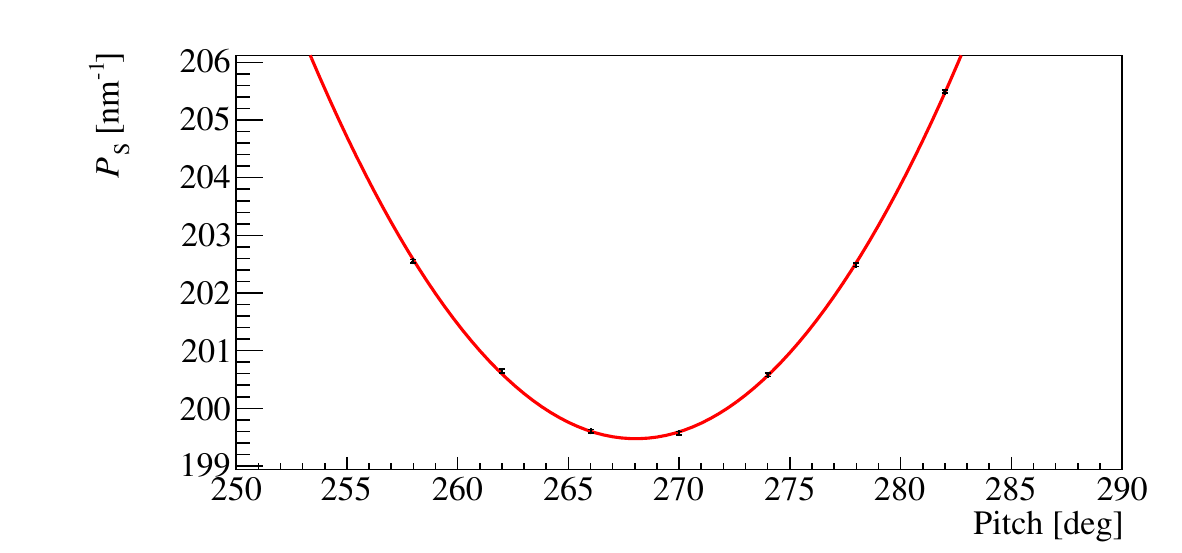}
    \caption{Measured $P_{\rm S}$ depending on the sample rotation of pitch.}
\label{fig:pitchalign_fringe}
\end{figure}
\begin{figure}[htbp]
  \centering
  \includegraphics[keepaspectratio,width=\linewidth]    {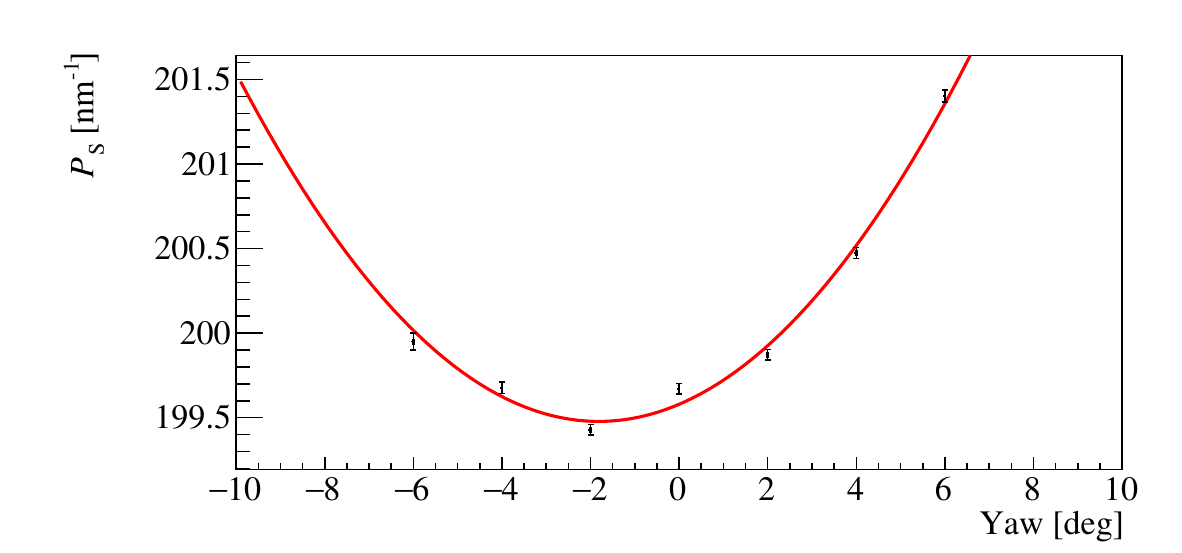}
        \caption{Measured $P_{\rm S}$ depending on the sample rotation of yaw.}
\label{fig:yawalign_fringe}
\end{figure}

The offset in the sample's rotation angle from perpendicularity to the beam axis changes the sample's effective thickness ($D_{\rm eff}$), as described  by the following equation:
\begin{equation}
    D_\mathrm{eff} = D/\cos\theta \simeq \frac{D}{1-\theta^2/2}, 
\end{equation}
where the $D$ is the sample thickness measured by the micrometer and $\theta$ is the sample rotation angle with respect to the beam axis.
To manage these misalignments, we measured the phase shifts of the samples in response to rotations around both the yaw and pitch axes. 
The phase shifts of a 1~mm-thick Al sample with respect to yaw and pitch are shown in Fig.~\ref{fig:pitchalign_fringe} and \ref{fig:yawalign_fringe}, respectively. 
For each sample, we fitted the measured phase shifts using a quadratic function and adjusted the orientation of the sample to minimize the phase shift.


For samples of V and V-Ni alloy, due to their smaller scattering lengths, did not allow observation of phase shifts induced by sample rotation. 
Therefore, these samples were visually oriented to be perpendicular to the beam axis. 
The uncertainty of rotation for the V and V-Ni alloy samples was determined using the deviation between the initial and minimal values obtained during the rotational adjustments of the Ti, Al, and Si samples. 
The resulting standard deviations for the yaw and pitch were $\sigma_\mathrm{pitch}=0.43~\unit{degrees}$ and $\sigma_\mathrm{yaw}=1.56~\unit{degrees}$, respectively. 
Notably, the uncertainty in the effective sample thickness due to rotation was significantly less than the uncertainty in the absolute thickness.


The neutron wavelength was determined using the TOF method with the distance from the moderator to detector $L$.
The $L$ was $17.74~\unit{m}$, and the size of the moderator was approximately 10~cm. 
Assuming that the value of $L $ only carries uncertainty equivalent to the size of the moderator, we adopted an uncertainty of $0.54\%$.


The atomic density $N$ was derived from the mass density, atomic weight, and Avogadro's constant. 
We used a specific gravity meter to determine the mass density of each sample to evaluate the effect of impurities, particularly in the V-Ni alloy. 
Although impurities were primarily suspected in the V and V-Ni alloys, we performed mass density measurements on all samples used in the experiments to ensure methodological consistency within this study.
The instrumental uncertainty was $0.082\%$, which was the primary uncertainty of the uncertainty of $N$.
The atomic weights are used in ref.\cite{Prohaska2022-jx}, and Avogadro's constant was $6.02214076\times 10^{23}~\unit{mol^{-1}}$.


Inserting a sample into one path of the interferometer reduces the atmospheric path by the thickness of the sample. The phase shift attributable to the sample was obtained by subtracting the atmospheric effect from the measured phase shift. The atmospheric phase shift was calculated using the following equation\cite{Rauch2015-ai}:
\begin{equation}
\phi_\mathrm{air} = (0.420\pm0.01~\unit{[m^{-12}]})D\lambda,
\end{equation}
where the $D$ is the sample thickness.
The factor of phase shift was calculated using an atmospheric pressure of $101.05~\unit{kPa}$, a temperature of $23\degC$, and a humidity of $15.25\%$.
Each of these values is an average over the experimental period.


Finally, the relative uncertainties of the neutron-nuclear scattering lengths measured by each sample are presented in table~\ref{table:uncertainty}. 
The values in the table indicate the relative uncertainty compared to the obtained $b_{\rm c}$.


\section{Impulity analysis}
\label{sec:purity}

We conducted an elemental analysis of the V and V-Ni alloy samples to explore the reason for the discrepancy in literature values. 
This analysis was performed using an X-ray fluorescence spectrometer (XRF) for elements ranging from Na to U, gas chromatography with thermal conductivity detection for H and N, and gas chromatography with non-dispersive infrared absorption methods for O. 
The XRF measurement was conducted by the RIKEN Materials Characterization Support Team, and the gas chromatography measurement was performed by TORAY Research Center, Inc. 
The samples used for XRF analysis were the same as those used in the experiment. 
The samples used for gas chromatography analysis were derived from the same base material as those used in the experiment. 
The measured contamination is documented in Table~\ref{table:contami_V}. 
Note that the $\rm V_a$ sample had a thickness of $0.316 ~\unit{mm}$ and the $\rm V_b$ sample was $0.314 ~\unit{mm}$ thick; both samples were cut from the same base material. 
Any elements with mixing ratios measured by XRF that fell below the detection or quantitation limits were excluded. 
Uncertainties in the mass mixing ratios from XRF measurements were only derived from statistical uncertainty. 
Due to the scarcity of data, uncertainties for the mass mixing ratios of H, N, and O could not be calculated, so the lower limit of quantitation was adopted.


Although a high concentration of hydrogen impurities might have explained the observed negative deviation, the measured concentration was relatively low at only 3~ppm. 
As such, it becomes clear that the observed discrepancy cannot be solely attributed to hydrogen impurity.
Vanadium metal is prone to oxygen contamination during the refining process\cite{Yoshinaga2007-wy}. 
Therefore, it is surmised that this contamination occurred during that process. 
Assuming the accuracy of the $b_{\rm c}$ values obtained in this study, the estimated oxygen mixing ratio in the sample used in the NIST database is approximately 0.9\%. 
Lending support to this hypothesis, a prior report suggested that the Bragg peak persisted unless the V-Ni alloy contained a lower amount of Ni than expected from the NIST database\cite{Sakurai2017-sq}.


\begin{table*}[hbtp]
\caption{Uncertainty table for each sample.}
\label{table:uncertainty}
\centering
\begin{ruledtabular}
\begin{tabular}{c|cccccccc}
sample & $t$ (mm) & statistics & systematics & \begin{tabular}{c} sample \\ thickness \end{tabular} & \begin{tabular}{c} sample \\ rotaion \end{tabular}& wavelength & \begin{tabular}{c} atomic \\ density \end{tabular} & air scat. \\ \colrule
Si & $0.287$ & $9.2\times 10^{-5}$ & $6.7\times 10^{-3}$ & $3.8\times 10^{-3}$ & $5.3\times 10^{-4}$ & $5.4\times 10^{-3}$ & $8.2\times 10^{-4}$ & $4.9\times 10^{-5}$ \\
Ti & $0.100$ & $4.4\times 10^{-4}$ & $1.8\times 10^{-2}$ & $1.7\times 10^{-2}$ & $1.0\times 10^{-5}$ & $5.4\times 10^{-3}$ & $8.2\times 10^{-4}$ & $5.1\times 10^{-5}$ \\
 & $0.204$ & $2.0\times 10^{-4}$ & $1.9\times 10^{-2}$ & $1.8\times 10^{-2}$ & $4.0\times 10^{-4}$ & $5.4\times 10^{-3}$ & $8.2\times 10^{-4}$ & $5.2\times 10^{-5}$ \\
Al & $0.103$ & $6.0\times 10^{-4}$ & $1.5\times 10^{-2}$ & $1.4\times 10^{-2}$ & $4.9\times 10^{-5}$ & $5.4\times 10^{-3}$ & $8.2\times 10^{-4}$ & $4.9\times 10^{-5}$ \\
 & $0.287$ & $5.3\times 10^{-5}$ & $7.8\times 10^{-3}$ & $5.6\times 10^{-3}$ & $4.5\times 10^{-5}$ & $5.4\times 10^{-3}$ & $8.2\times 10^{-4}$ & $4.9\times 10^{-5}$ \\
 & $0.961$ & $4.1\times 10^{-5}$ & $5.7\times 10^{-3}$ & $1.5\times 10^{-3}$ & $7.5\times 10^{-6}$ & $5.4\times 10^{-3}$ & $8.2\times 10^{-4}$ & $4.8\times 10^{-5}$ \\
V & $0.316$ & $3.9\times 10^{-4}$ & $6.8\times 10^{-3}$ & $4.1\times 10^{-3}$ & $3.9\times 10^{-4}$ & $5.3\times 10^{-3}$ & $8.1\times 10^{-4}$ & $2.7\times 10^{-4}$ \\
 & $0.314$ & $6.1\times 10^{-4}$ & $7.2\times 10^{-3}$ & $4.7\times 10^{-3}$ & $3.9\times 10^{-4}$ & $5.3\times 10^{-3}$ & $8.1\times 10^{-4}$ & $2.7\times 10^{-4}$ \\
V-Ni & $0.315$ & $3.1\times 10^{-3}$ & $1.8\times 10^{-2}$ & $1.7\times 10^{-2}$ & $3.6\times 10^{-4}$ & $4.9\times 10^{-3}$ & $7.8\times 10^{-4}$ & $2.2\times 10^{-3}$ \\

\end{tabular}
\end{ruledtabular}
\end{table*}


\begin{table*}[htb]
\caption{Contamination of each sample. The unit of these values are wt\%}
\label{table:contami_V}
\centering
\begin{ruledtabular}
\begin{tabular}{c|ccccc}
atom & {$M$} & {$b_{\rm c}$ (fm)} & {V-Ni alloy (wt\%)} & {$\rm V_a$ (wt\%)} & {$\rm V_b$ (wt\%)}\\ \colrule
H & 1.01 & -3.739 $\pm$ 0.0011 & 0.0002 $\pm$ 0.0002 & 0.0003 $\pm$ 0.0002 & 0.0003 $\pm$ 0.0002 \\
N & 14.01 & 9.36 $\pm$ 0.02 & 0.02 $\pm$ 0.006 & 0.02 $\pm$ 0.006 & 0.02 $\pm$ 0.006 \\
O & 16.00 & 5.803 $\pm$ 0.004 & 0.023 $\pm$ 0.006 & 0.019 $\pm$ 0.006 & 0.019 $\pm$ 0.006 \\
Al & 26.98 & 3.449 $\pm$ 0.005 & 0.106 $\pm$ 0.0016 & 0.15 $\pm$ 0.0018 & 0.166 $\pm$ 0.002 \\
Si & 28.09 & 4.1491 $\pm$ 0.001 & 0.039 $\pm$ 0.00064 & 0.0666 $\pm$ 0.00078 & 0.0658 $\pm$ 0.0008 \\
P & 30.97 & 5.13 $\pm$ 0.01 & 0.013 $\pm$ 0.00021 & 0.0173 $\pm$ 0.00021 & 0.0145 $\pm$ 0.0006 \\
S & 32.06 & 2.847 $\pm$ 0.001 & 0.02 $\pm$ 0.00021 & 0.0229 $\pm$ 0.00021 & 0.0194 $\pm$ 0.0002 \\
V & 50.94 & -0.3824 $\pm$ 0.0012 & 94.7 & 99.7 & 99.7 \\
Fe & 55.84 & 9.45 $\pm$ 0.02 & 0.029 $\pm$ 0.0012 & 0.0119 $\pm$ 0.00081 & 0.033 $\pm$ 0.0002 \\
Ni & 58.69 & 10.3 $\pm$ 0.1 & 5.055 $\pm$ 0.0096 & 0.0036 $\pm$ 0.0005 & {      ---      } \\
Cu & 63.55 & 7.718 $\pm$ 0.004 & {      ---      } & 0.00745 $\pm$ 0.00035 & 0.0057 $\pm$ 0.0002 \\
Zn & 65.38 & 5.68 $\pm$ 0.005 & 0.0088 $\pm$ 0.00035 & 0.00444 $\pm$ 0.00024 & 0.0039 $\pm$ 0.0005 \\
Ge & 72.63 & 8.185 $\pm$ 0.02 & 0.014 $\pm$ 0.0014 & 0.014 $\pm$ 0.0011 & 0.0135 $\pm$ 0.0003 \\
As & 74.92 & 6.58 $\pm$ 0.01 & {      ---      } & {      ---      } & 0.0019 $\pm$ 0.0004 \\
Sn & 118.71 & 6.225 $\pm$ 0.002 & 0.0081 $\pm$ 0.0006 & 0.006 $\pm$ 0.0006 & {      ---      } \\
\end{tabular}
\end{ruledtabular}
\end{table*}



\clearpage
\bibliography{ref}

\end{document}